\documentclass{PoS}

\title{SKA-VLBI Key Science Programmes}

\ShortTitle{SKA-VLBI KSPs}

\author{\speaker{Zsolt Paragi}\thanks{The \lq\lq VLBI with the SKA'' work package is part of the JUMPING JIVE project, that
has received funding from the European Union's Horizon 2020 research and innovation programme under grant agreement No 730884.}\\
        Joint Institute for VLBI ERIC (JIVE), \\ Oude Hoogeveensedijk 4, 7991\,PD Dwingeloo, The Netherlands\\
        E-mail: \email{zparagi@jive.eu}}
       
\author{Antonio Chrysostomou,  Cristina Garc\'ia-Mir\'o\\
        Square Kilometre Array Organisation (SKA), Jodrell Bank Observatory,\\ 
        Lower Withington, Macclesfield, Cheshire SK11 9DL, United Kingdom\\
        E-mail: \email{a.chrysostomou@skatelescope.org, c.garcia-miro@skatelescope.org}}

\abstract{A significant fraction of the observing time with the two phase-I SKA telescopes (SKA1-LOW \& SKA1-MID) will be spent on Key Science Projects led by member country scientists. The various SKA Science Working Groups, including the VLBI Focus Group are in the process of defining KSPs that are aligned with the High Priority Science Objectives of the SKA. At the moment it is not clear how the special observing mode of SKA-VLBI - when the SKA1 components are phased-up and included in VLBI networks - could be incorporated in KSPs. The VLBI community needs to be prepared by the time the KSP proposal calls are expected (mid-2020s). In this paper we outline the basic concept of SKA-VLBI, and some possibilities for us to engage in SKA KSPs.}

\FullConference{14th European VLBI Network Symposium \& Users Meeting (EVN 2018)\\
		8-11 October 2018\\
		Granada, Spain}

\begin{document}

\section{VLBI with the SKA}

The Square Kilometre Array (SKA) is a next generation radio facility that will eventually have a collecting area of about a square kilometre, and baselines up to thousands of kilometres\footnote{http://www.skatelescope.org}. The science case of this ultimate very long baseline interferometer (VLBI) array was initially published at the end of the 1990s \cite{SKA99}, and the first comprehensive science case was compiled in 2004 \cite{SKA04}. The challenge of building a fast survey machine that images large portion of the sky in a relatively short time, but at very high resolution, was realised from the very beginning. The \lq\lq Carilli \& Rawlings'' book in 2004 highlighted five possible key science projects. These were \lq\lq The cradle of life'', \lq\lq Strong-field tests of gravity using pulsars and black holes'', \lq\lq The origin and evolution of cosmic magnetism'', \lq\lq Galaxy evolution, cosmology and dark energy'', and \lq\lq Probing the dark ages''. With time it became clear that the various envisaged imaging surveys (EOR, continuum, H{\sc i} etc.) and time domain surveys (pulsar search, pulsar timing) require most of the collecting area in very short spacings, and this has driven the design of the instrument ever since. There will be two telescopes of phase I SKA built with different frequency coverages. SKA1-LOW (50\,MHz -- 350\,MHz) will be in Australia, with maximum baseline length of about 65\,km. SKA1-MID (0.35\,GHz -- 15.3(24.0)\,GHz) will be built in South Africa, with maximum baseline length of about 150\,km. The second phase of the SKA will provide the full sensitivity, and a resolution that is 20 times that of the phase I components.  The latest collection of science cases were published in 2015 by the SKA Organisation (SKAO), with headquarters in Jodrell Bank, UK \cite{SKA15}.

\begin{figure}[!ht]
\vspace{1.0 cm}
\includegraphics[angle=0,width=6.2in]{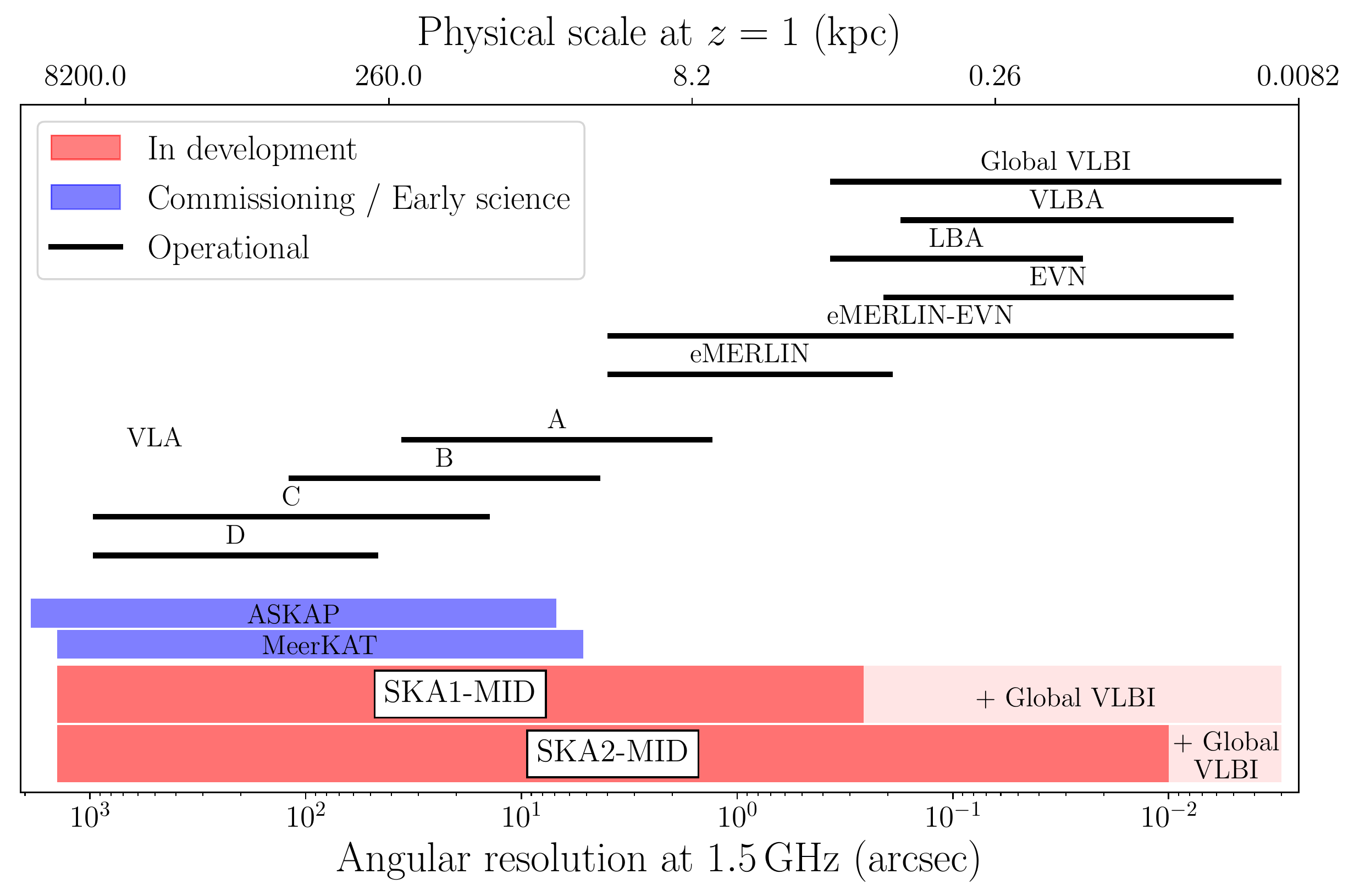}
\includegraphics[angle=0,width=6.0in]{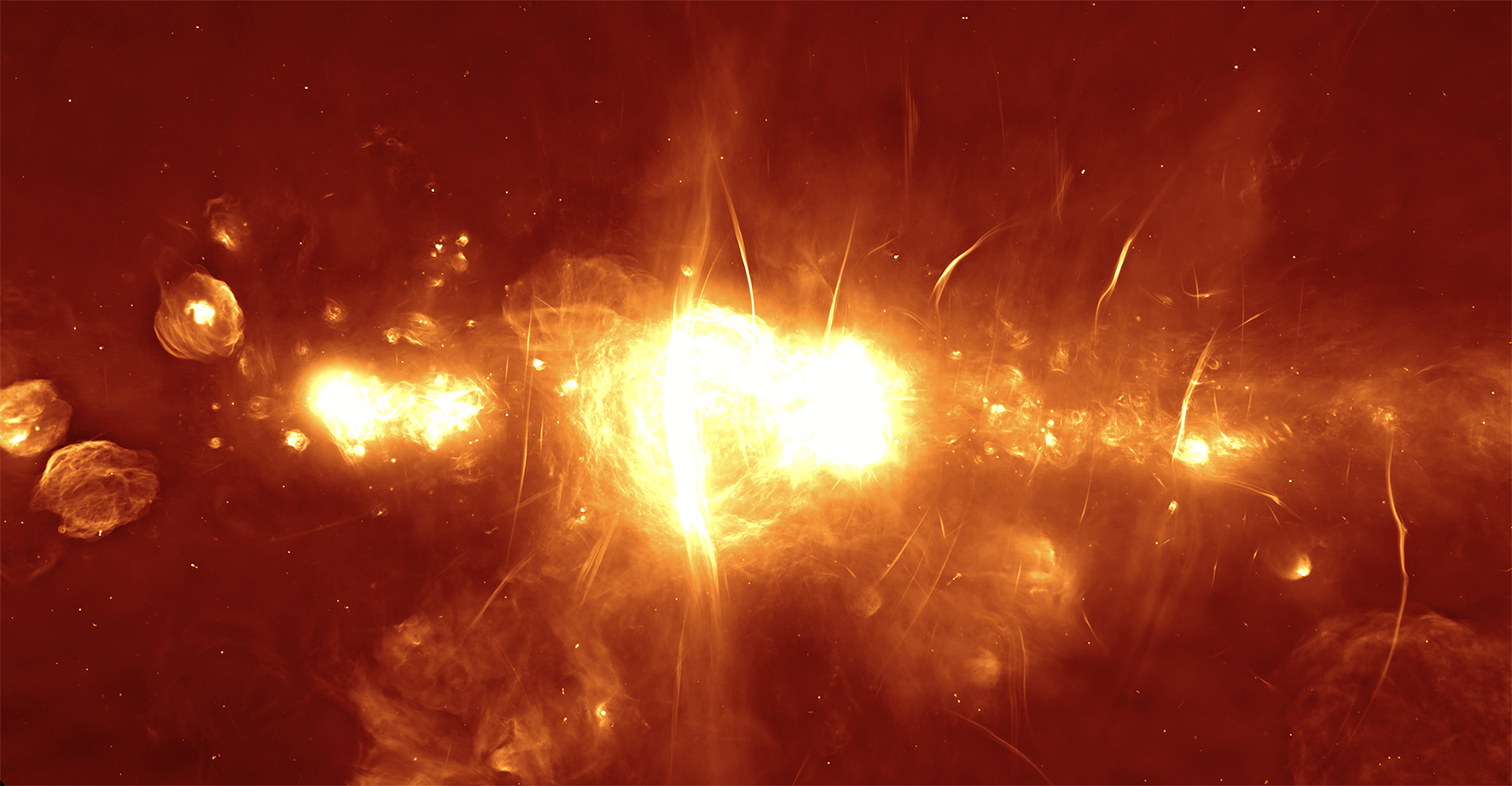} 
\caption{Above: angular scales (and corresponding spatial scales at $z=1$) probed by various current and future interferometers working in the cm wavebands. Courtesy of Jack Radcliffe. Below: the Galactic Centre -- the first public image from the 64-element SKA1-MID precursor MeerKAT
(Courtesy of SARAO).}
\label{fig1}
\end{figure}

The SKAO is in continuous discussion with the community about how to best address its science goals. There are a number of Science Working Groups and Focus Groups working on this. A main concern for the VLBI community is how to carry out very high angular resolution science during phase I of the project. In 2015 the VLBI working group was formally established, with co-chairs Cormac Reynolds (CSIRO) and Zsolt Paragi (JIVE)-- the latter replaced in 2018 by Tao An (ShAO). According to plans, the core of the SKA phase I components will be phased-up. In the past this was the standard operation at the Westerbork Synthesis Radio Telescope during European VLBI Network (EVN) observations, and it is routinely done for VLBI observations with ALMA as well. A single tied-array beam will be significantly smaller than an arcsecond at GHz frequencies, but it will be possible to form several independent beams in the same sub-array, looking at different targets/calibrators within the primary field of view of the SKA dishes \cite{SKA-VLBI15,SKA-VLBI18}. The VLBI working group helped to define L0 science requirements, and L1 technical requirements together with the various SKA Consortia, and established a few initial use cases. One of the important requirements was for simultaneous VLBI (phased-array) and SKA1 interferometer data products. The advantages are straightforward for calibration: in VLBI there are no primary flux density calibrators, because compact sources are variable. 
The SKA data will provide a way to measure the flux densities of compact sources accurately. 
For science: we will have for the first time simultaneous information on a wide range of angular scales, clearly beneficial to understand our targets together with their environments, from arcminute to (sub-)milliarcsecond scales. This is demonstrated in Fig.~\ref{fig1}. Another noteworthy requirement is real-time VLBI data streaming capability from the SKA1 components: the e-EVN is an SKA pathfinder not just as an instrument, but it can also be considered as a pathfinder for SKA-VLBI operations. 

\section{EVN and SKA-VLBI science priorities}

\begin{centering}
\begin{figure}[!ht]
\vspace{1.0 cm}
\includegraphics[angle=0,width=5.4in]{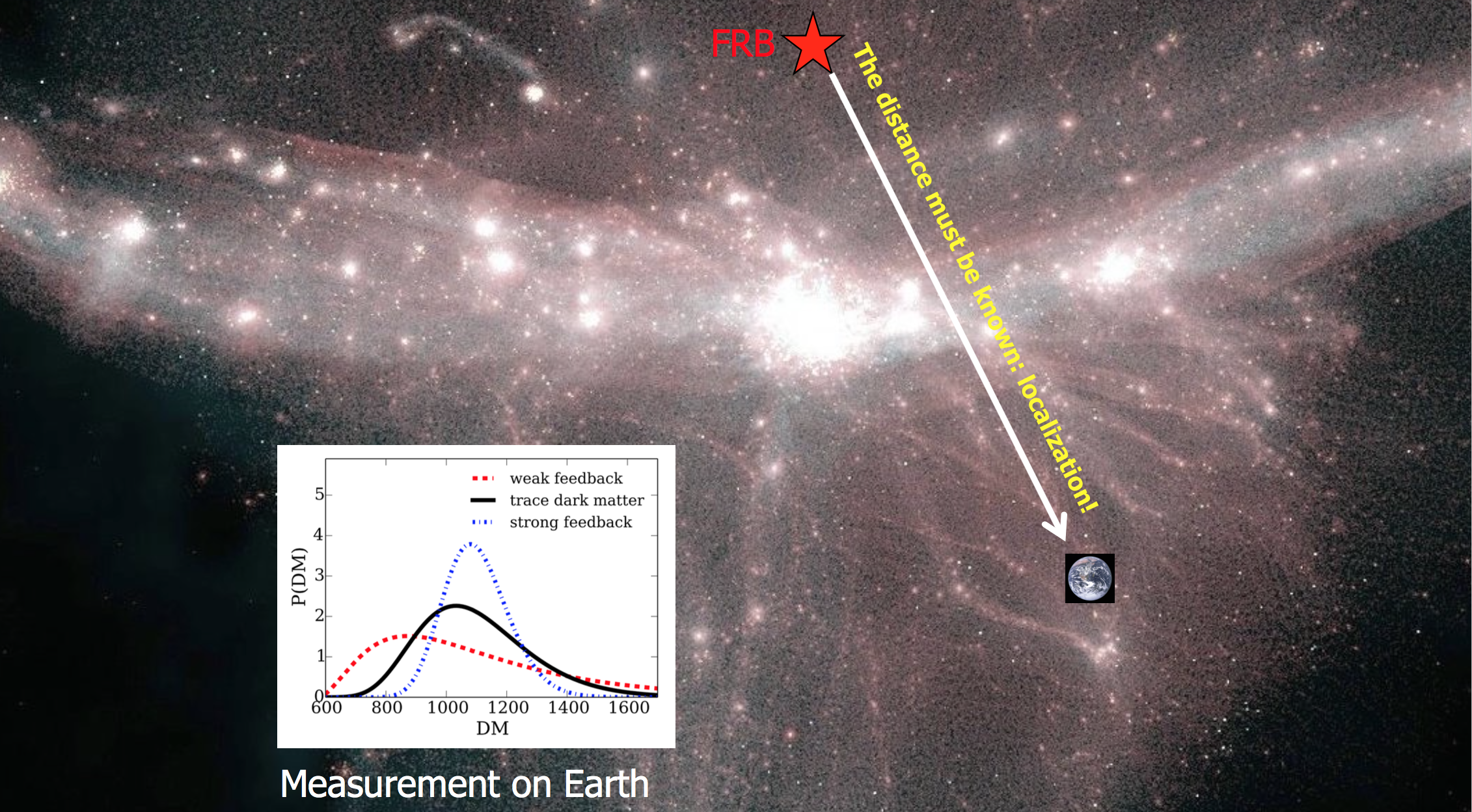} 
\includegraphics[angle=0,width=5.4in]{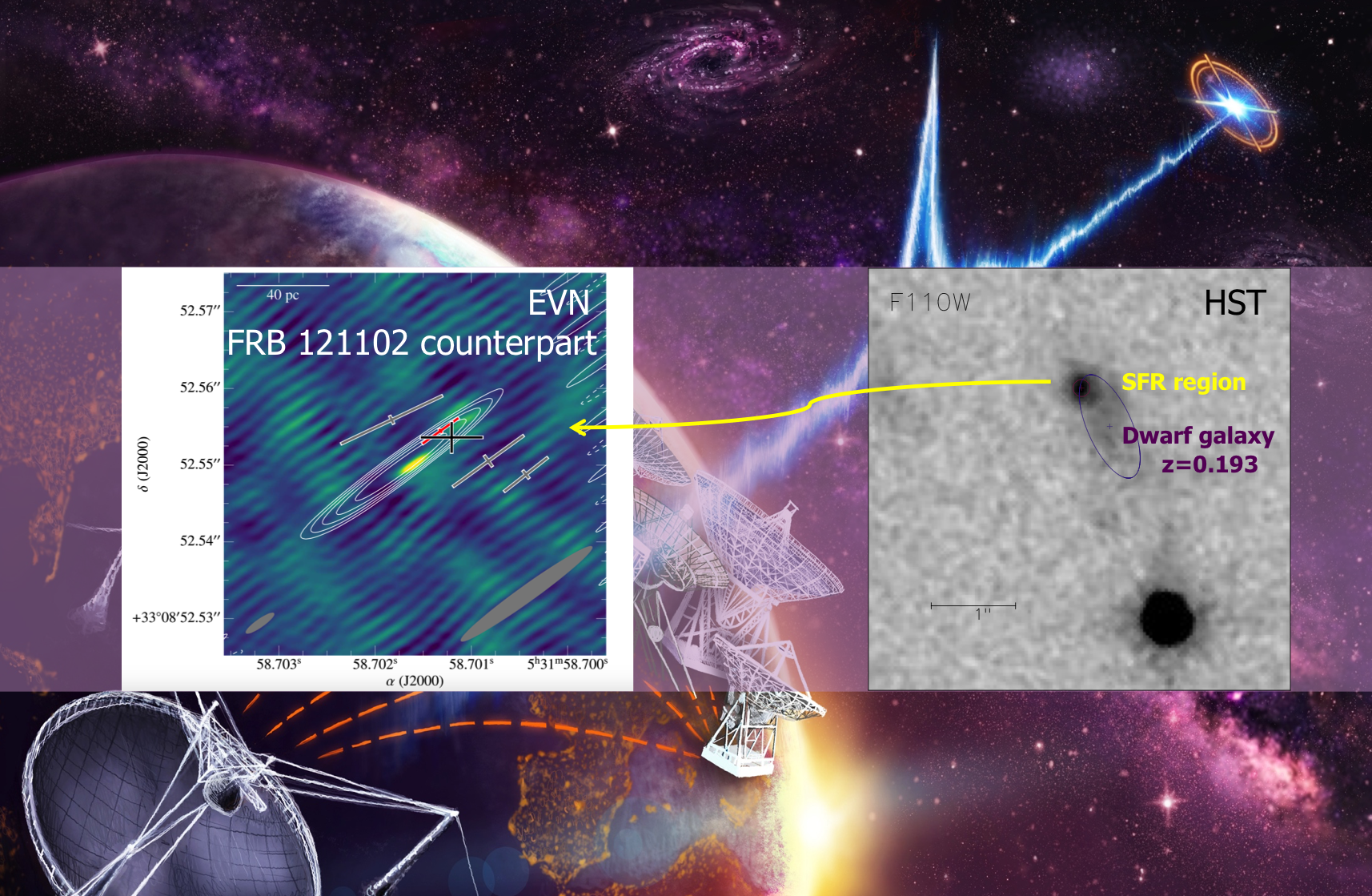} 
\caption{Above: one of the highest priority science objectives of the SKA. The milliseconds-duration signals from fast radio bursts (FRBs) get dispersed in the intergalactic medium. The distribution of the measured dispersion measures (DM; inset from \cite{DM}) for several hundreds of FRBs -- in a given redshift bin -- sheds light on the distribution of baryonic matter in the Universe. This requires sub-arcsec localization at high redshifts. Background image: {\em https://www.alanrduffy.com}. Below: JVLA and EVN localization of the repeating FRB121102. The short pulses detected by the EVN are co-located within $\sim$10~mas of a compact, persistent radio source within a star forming region in a dwarf galaxy at $z=0.1927$. Ultimate evidence that FRB-like signals (may) have cosmological origin \cite{JVLAFRB,EVNFRB,OPTFRB,HSTFRB}. Artist's impression: Danielle Futselaar.}
\label{fig2}
\end{figure}
\end{centering}

There were a number of high profile VLBI results that motivated us to push for VLBI capabilities in the SKA \cite{SKA-VLBI15}. For example the accurate distance determination of the dwarf nova SS~Cyg -- requiring astrometric VLBI observations triggered within a day -- that resolved a debate about accretion disc theory \cite{SSCyg}, the direct evidence for jet-driven, large-scale neutral and molecular outflow in the recurrent young radio source 4C12.50 \cite{4C12.50}, and the identification of shocks where very high energy emission originates in the classical Nova Monoceros 2012 \cite{NovaMon}. Large field-of-view VLBI surveys \cite{Chi}, comparing global VLBI astrometry results with future Gaia catalogs, and observing tidal disruption events (TDE) were in the wish list for ultra-sensitive SKA-VLBI observations as well \cite{SKA-VLBI15}. It is still somewhat a surprise just how much our field has developed in the past three years. We have detected and localised a repeating fast radio burst (FRB) on milliarcsecond scales, which opened a whole new chapter in the VLBI study of short-timescale transients (see Fig.~\ref{fig2}, and references). We can measure the jet collimation profile in an active galactic nucleus (AGN) from $10^2$ to $10^4$ gravitational radii from the supermassive black hole, using space-Earth baselines to RadioAstron \cite{3C84}. We detected dozens of targets -- also aided by primary beam correction within the EVN -- in deep fields \cite{HDF}, we found evidence for kpc-scale optical jets by comparing VLBI and Gaia astrometry \cite{VLBI-Gaia}, and we resolved ejecta from a TDE in Arp~299B \cite{Arp299B}. Perhaps most unexpected, we resolved ejecta of the first-ever electromagnetic counterpart to a gravitational wave event GW1701817 \cite{GW170817-HSA,GW170817-global}. All of these results have been presented at this symposium. Clearly, there is a lot of momentum to push for major VLBI observing programmes on these (and other) topics, that include the SKA.

The range of topics largely overlaps with the science the EVN pursues in the coming decade \cite{EVN-Future}. We have to identify those areas where the SKA1 components will have a unique contribution to VLBI. For example: surprising as it may sound, the quality of amplitude calibration of an interferometer has an effect on its resolution power in the high signal-to-noise ratio regime ($\sim$S/N>100; \cite{Natarajan}). Station calibration at the level of $\sim1\%$ will be needed to accurately measure source sizes (under some assumptions) smaller than about one tenth of the beam size. While at low S/N knowing the total flux density is critical (cf. \cite{GW170817-global}) -- SKA1 data on targets and calibrators will help in both cases. Similarly, SKA1 will provide very precise polarization characterisation of our targets and calibrators on arcsec scales, which will help to accurately calibrate the cross-hand VLBI data products, to measure the frequency dependent fractional linear and circular polarizations, the polarization angles as well as the rotation measure in source components seen on VLBI scales. The multi-beam capability of SKA1 telescopes means that it will be possible to use a number of calibrators for ultra-precise (down to a few microarcseconds) relative astrometry, on sub-mJy targets. This has important implications for stellar (continuum and maser), pulsar, and transient astrometry observations. For example, maser astrometry with the SKA is a promising tool to measure the size and rotation of the Galaxy \cite{MW-simulations}. There is a lot of synergy in VLBI studies of various SKA1 survey fields or particular targets of interest, let it be continuum, spectral line, or time-domain, even if the SKA1 components are not always part of the follow-up VLBI observations. The demand on VLBI observations will likely increase to an extent that requires an extension to current observing sessions, possibly also allowing for more regular \lq\lq EVN-lite" observations with a subset of telescopes, to follow-up SKA1 triggers (an idea that has been around in the EVN for a while).  

\section{SKA Key Science Programmes: strategies for VLBI}

The SKAO and the science working groups have compiled a list of forty high priority science objectives from eight broad topics, and a subset of thirteen \lq\lq highest-ranked science objectives''.
These range from studies of the epoch of reionisation with SKA1-LOW, test of general relativity with pulsars (MID), observing atomic hydrogen nearby and at high redshifts (MID/LOW), FRB cosmology (MID/LOW), \lq\lq Cradle of Life'' (planet and star-formation, structure of the Milky Way using SKA1-MID), the origin of magnetic fields (MID/LOW), and continuum surveys for cosmology as well as for the study of star formation history in the Universe (MID). Several of these will heavily rely on additional VLBI observations. There have been initial SKA-VLBI \lq\lq use cases'' established for pulsar astrometry, stellar deep fields, Galactic Structure (two independent maser use cases), synchrotron transient follow-up,  and AGN surveys. These use cases aim to define the observations and the required SKA resources for a given science goal. More of these ideas will have to be worked out in more detail, as the SKA1 design stabilises through 2018/2019. 

The KSPs for the SKA have not been defined yet, but these will be large projects \lq\lq that require large observing time allocations over a period longer than one Time allocation cycle'' (SKA KSP Framework Paper). It is expected that a typical KSP will ask for 1000h of observations over 3--4 years. This is about the same as having for example all ten 24h e-EVN sessions per year dedicated fully to a single SKA-VLBI KSP for four years. Having matched resources might be a challenge, but may not be that critical considering the global VLBI resources available world-wide. The bottle neck might actually be the time available for VLBI at the SKA1 components. Therefore it is not only the science cases, but their possible realization will have to be thought through carefully. One way of getting around this problem is to incorporate a SKA-VLBI element (requiring $<\!<1000$\,h in itself) in regular SKA KSPs. Another possibility would be doing piggy-back VLBI observations on various SKA1 surveys, although this may require some special VLBI scheduling requirements to be taken into account when organizing the survey observations. We will have to understand the limitations of commensal operations within the SKA very well. This is important also because we will preferentially need the SKA1 data products in all regular SKA-VLBI observations. Another consideration is to request some of the SKA1 dishes -- those that are usually not used in SKA surveys requiring dishes in the core of the array only -- being available for VLBI for extended periods of time (for sufficiently bright targets). 

These ideas will be discussed in specific SKA and SKA-VLBI Key Science Projects workshops, starting in 2019. While the final construction of the SKA1 components is in the future, it is worth noting that MeerKAT is already operational (see Fig.~\ref{fig1}), and the first dish of the future African VLBI Network (the Ghana telescope) has already produced VLBI fringes. We may also expect the SKAO will allow for shared-risk science verification observations during commissioning.  It is time we get ready for SKA-VLBI!


\begin{thebibliography}{99}
\bibitem{SKA99} R. Taylor, R. Braun (eds.), {\it Science with the Square Kilometre Array, A Next Generation World Radio Observatory}, Netherlands Foundation for Radio Astronomy, Dwingeloo, 1999
\bibitem{SKA04} C. Carilli, S. Rawlings (eds.), {\it Science with the Square Kilometre Array}, Elsevier B.V., Amsterdam, 2004
\bibitem{SKA15} T. Bourke, R. Braun, R. Fender et al. (eds.), {\it Advancing Astrophysics with the Square Kilometre Array}, Dolman Scott Ltd, SKAO, 2015;  Published online as 
\href{https://pos.sissa.it/cgi-bin/reader/conf.cgi?confid=215}{PoS[AASKA14]}
\bibitem{SKA-VLBI15} Z. Paragi, L. Godfrey, C. Reynolds et al., {\it Very Long Baseline Interferometry with the SKA}, {\it Proceedings of Science}, 
\href{https://pos.sissa.it/215/143/pdf}{PoS[AASKA14]143}, 2015 
\bibitem{SKA-VLBI18} C. Garc\'ia-Mir\'o, {\it The science impact of high sensitivity VLBI with SKA}, {\it these proceedings}
\bibitem{DM} J.\,P. Macquart, E. Keane, K. Grainge et al., {\it Fast Transients at Cosmological Distances with the SKA}, {\it Proceedings of Science}, \href{https://pos.sissa.it/215/55/pdf}{PoS[AASKA14]55}, 2015
\bibitem{JVLAFRB} S. Chatterjee, C.\,J. Law, R.\,S. Wharton et al., {\it A direct localization of a fast radio burst and its host}, {\it Nature}, {\bf 541} (2017) 58
\bibitem{EVNFRB} B. Marcote, Z. Paragi, J.\,W.\,T. Hessels et al., {\it The Repeating Fast Radio Burst FRB 121102 as Seen on Milliarcsecond Angular Scales}, {\it ApJ}, {\bf 834} (2017) L8
\bibitem{OPTFRB} S.\,P. Tendulkar, C.\,G. Bassa, J.\,M. Cordes et al., {\it The Host Galaxy and Redshift of the Repeating Fast Radio Burst FRB 121102}, {\it ApJ}, {\bf 834} (2017) L7
\bibitem{HSTFRB} C.\,G. Bassa, S.\,P. Tendulkar, E.\,A.\,K. Adams et al., {\it FRB 121102 Is Coincident with a Star-forming Region in Its Host Galaxy}, {\it ApJ}, {\bf 843} (2017) L8 
\bibitem{SSCyg} J.\,C.\,A. Miller-Jones, G.\,R. Sivakoff, C. Knigge et al., {\it An Accurate Geometric Distance to the Compact Binary SS Cygni Vindicates Accretion Disc Theory}, {\it Science}, {\bf 340} (2013) 950
\bibitem{4C12.50} R. Morganti, J. Fogasy, Z. Paragi, T. Oosterloo, M. Orienti, {\it Radio Jets Clearing the Way Through a Galaxy: Watching Feedback in Action}, {\it Science}, {\bf 341} (2013) 1082
\bibitem{NovaMon} L. Chomiuk, J.\,D. Linford, Yang, J. et al., {\it Binary orbits as the driver of $\gamma$-ray emission and mass ejection in classical novae}, {\it Nature}, {\bf 514} (2014) 339
\bibitem{Chi} S. Chi, P.\,D. Barthel, M.\,A. Garrett, {\it Deep, wide-field, global VLBI observations of the Hubble deep field north (HDF-N) and flanking fields (HFF)}, {\it A\&A}, {\bf 550} (2013) A68
\bibitem{3C84} G. Giovannini, T. Savolainen, M. Orienti et al., {\it A wide and collimated radio jet in 3C84 on the scale of a few hundred gravitational radii}, {\it Nature Astronomy}, {\bf 2} (2018) 472
\bibitem{HDF} J.\,F. Radcliffe, M.\,A. Garrett, T.\,W.\,B. Muxlow et al., {\it Nowhere to Hide: Radio-faint AGN in GOODS-N field. I. Initial catalogue and radio properties}, {\it A\&A}, {\bf 619} (2018) A48
\bibitem{VLBI-Gaia} Y.\,Y. Kovalev, L. Petrov, A. Plavin, {\it VLBI-Gaia offsets favor parsec-scale jet direction in active galactic nuclei}, {\it A\&A}, {\bf 598} (2017) L1 
\bibitem{Arp299B} S. Mattila, M.\,A. P\'erez-Torres, A. Efstathiou, {\it A dust-enshrouded tidal disruption event with a resolved radio jet in a galaxy merger}, {\it Science}, {\bf 361} (2018) 482 
\bibitem{GW170817-HSA} K.\,P. Mooley, A.\,T. Deller, O. Gottlieb, {\it Superluminal motion of a relativistic jet in the neutron-star merger GW170817}, {\it Nature}, {\bf 561} (2018) 355
\bibitem{GW170817-global} G. Ghirlanda, O. Salafia, Z. Paragi et al., {\it Re-solving the jet/cocoon riddle of the first gravitational wave with an electromagnetic counterpart} [\href{http://de.arxiv.org/abs/1808.00469}{\tt arXiv:1808.00469}]
\bibitem{EVN-Future} T. Venturi, M. Lindqvist, Z. Paragi (eds.) {\it The Future of the EVN} (in prep.)
\bibitem{Natarajan} I. Natarajan, Z. Paragi, J. Zwart et al., {\it Resolving the blazar CGRaBS J0809+5341 in the presence of telescope systematics}, {\it MNRAS}, {\bf 464} (2017) 4306
\bibitem{MW-simulations} L.~H. Quiroga-Nu{\~n}ez, H.~J. van Langevelde, M.~J. Reid, J.~A. Green,  {\it Simulated Galactic methanol maser distribution to constrain Milky Way parameters}, {\it A\&A}, {\bf 604} (2017) A72


\end{thebibliography}
\end{document}